\documentclass{elsart}

\usepackage{graphicx}

\newcommand{\KK}{\ensuremath{\rm{K}^+\rm{K}^-}}

\newcommand{\mmd}{\ensuremath{\mbox{MM}_d(\gamma,\rm{KK})}}
\newcommand{\Eg}{\ensuremath{{E}_{\gamma}}~}

\begin{document}

\begin{frontmatter}

\title{Forward coherent $\phi$-meson photoproduction from deuterons
  near threshold}

%%%%%%%%%%%%%%%%%%%%%%%%%%%%%%%%%%%
%\include{LEPS_collab_plb.tex}
%%%%%%%%%%%%%%%%%%%%%%%%%%%%%%%%%%%
%%%%%%%%%%%%%%%%%%%%%%%%% Collaborators %%%%%%%%%%%%%%%%%%%%%%%%%%
\author{W.C.~Chang,$^{a}$}
\author{K.~Horie,$^{b}$}
\author{S.~Shimizu,$^{b}$}
\author{M.~Miyabe,$^{c}$}
\author{D.S.~Ahn,$^{b,d}$}
\author{J.K.~Ahn,$^{d}$}
\author{H.~Akimune,$^{e}$}
\author{Y.~Asano,$^{f}$}
\author{S.~Dat\'e,$^{g}$}
\author{H.~Ejiri,$^{b,g}$}
\author{S.~Fukui,$^{h}$}
\author{H.~Fujimura,$^{i,c}$}
\author{M.~Fujiwara,$^{b,f}$}
\author{S.~Hasegawa,$^{b}$}
\author{K.~Hicks,$^{j}$}
\author{T.~Hotta,$^{b}$}
\author{K.~Imai,$^{c}$}
\author{T.~Ishikawa,$^{k}$}
\author{T.~Iwata,$^{l}$}
\author{Y.~Kato,$^{b}$} 
\author{H.~Kawai,$^{m}$}
\author{Z.Y.~Kim,$^{i}$}
\author{K.~Kino,$^{b}$}
\author{H.~Kohri,$^{b}$}
\author{N.~Kumagai,$^{g}$}
\author{P.J.~Lin,$^{n}$} 
\author{S.~Makino,$^{o}$}
\author{T.~Matsuda,$^{p}$}
\author{T.~Matsumura,$^{q}$}
\author{N.~Matsuoka,$^{b}$}
\author{T.~Mibe,$^{j}$}
\author{Y.~Miyachi,$^{r}$}
\author{M.~Morita,$^{b}$}
\author{N.~Muramatsu,$^{f,b}$}
\author{T.~Nakano,$^{b}$}
\author{M.~Niiyama,$^{c}$}
\author{M.~Nomachi,$^{s}$}
\author{Y.~Ohashi,$^{g}$}
\author{H.~Ohkuma,$^{g}$}
\author{T.~Ooba,$^{m}$}
\author{D.S.~Oshuev,$^{a}$}
\author{C.~Rangacharyulu,$^{t}$}
\author{A.~Sakaguchi,$^{s}$}
\author{T.~Sasaki,$^{c}$}
\author{P.M.~Shagin,$^{u}$}
\author{Y.~Shiino,$^{m}$}
\author{A.~Shimizu,$^{b}$}
\author{H.~Shimizu,$^{k}$}
\author{Y.~Sugaya,$^{s}$}
\author{M.~Sumihama,$^{b,f}$}
\author{Y.~Toi,$^{p}$}
\author{H.~Toyokawa,$^{g}$}
\author{A.~Wakai,$^{v}$}
\author{C.W.~Wang,$^{a}$}
\author{S.C.~Wang,$^{a}$}
\author{K.~Yonehara,$^{e}$}
\author{T.~Yorita,$^{b,g}$}
\author{M.~Yoshimura,$^{w}$}
\author{M.~Yosoi,$^{c,b}$}
\author{and R.G.T.~Zegers,$^{x}$} 
\author{(LEPS Collaboration)}
\address[a]{Institute of Physics, Academia Sinica, Taipei, Taiwan 11529, Taiwan}
\address[b]{Research Center for Nuclear Physics, Osaka University, Ibaraki, Osaka 567-0047, Japan}
\address[c]{Department of Physics, Kyoto University, Kyoto 606-8502, Japan} 
\address[d]{Department of Physics, Pusan National University, Busan 609-735, Korea}
\address[e]{Department of Physics, Konan University, Kobe, Hyogo 658-8501, Japan}
\address[f]{Kansai Photon Science Institute, Japan Atomic Energy Agency, Kizu, Kyoto 619-0215, Japan}
\address[g]{Japan Synchrotron Radiation Research Institute, Mikazuki, Hyogo 679-5198, Japan}
\address[h]{Department of Physics and Astrophysics, Nagoya University, Nagoya, Aichi 464-8602, Japan}
\address[i]{School of Physics, Seoul National University, Seoul, 151-747, Korea}
\address[j]{Department of Physics And Astronomy, Ohio University, Athens, Ohio 45701, USA}
\address[k]{Laboratory of Nuclear Science, Tohoku University, Sendai, Miyagi 982-0826, Japan}
\address[l]{Department of Physics, Yamagata University, Yamagata 990-8560, Japan}
\address[m]{Department of Physics, Chiba University, Chiba 263-8522, Japan}
\address[n]{Department of Physics, National Kaohsiung Normal University, Kaohsiung 824, Taiwan}
\address[o]{Wakayama Medical College, Wakayama, Wakayama 641-8509, Japan}
\address[p]{Department of Applied Physics, Miyazaki University, Miyazaki 889-2192, Japan}
\address[q]{Department of Applied Physics, National Defense Academy, Yokosuka 239-8686, Japan}
\address[r]{Department of Physics, Tokyo Institute of Technology, Tokyo 152-8551, Japan} 
\address[s]{Department of Physics, Osaka University, Toyonaka, Osaka 560-0043, Japan}
\address[t]{Department of Physics, University of Saskatchewan, Saskatoon, Saskatchewan, Canada} 
\address[u]{School of Physics and Astronomy, University of Minnesota, Minneapolis, Minnesota 55455, USA}
\address[v]{Akita Research Institute of Brain and Blood Vessels, Akita 010-0874, Japan}
\address[w]{Institute for Protein Research, Osaka University, Suita, Osaka 565-0871, Japan}
\address[x]{National Superconducting Cyclotron Laboratory, Department of Physics and Astronomy, Michigan State University, East Lansing, MI 48824-1321, USA}
%%%%%%%%%%%%%%%%%%%%%%%%%%%%%%%%%%%%%%%%%%%%%%%%%%%%%%%%%%%%%%%%%%%%%%%%

\begin{abstract}

Differential cross sections and decay asymmetries for coherent
$\phi$-meson photoproduction from deuterons were measured for the
first time at forward angles using linearly polarized photons at
$E_{\gamma}$= 1.5-2.4 GeV. This reaction offers a unique way to
directly access natural-parity Pomeron dynamics and gluon exchange at
low energies. The cross sections at zero degrees increase with
increasing photon energy. The decay asymmetries demonstrate a complete
dominance of natural-parity exchange processes, showing that isovector
unnatural-parity $\pi$-meson exchange is small. Nevertheless the
deduced cross sections of $\phi$-mesons from nucleons contributed by
isoscalar t-channel exchange processes are not well described by the
conventional Pomeron model.

\end{abstract}


\end{frontmatter}

% main text
\section{Introduction}
The common asymptotic behavior of high-energy diffractive processes of
hadron-hadron and photon-hadron interactions is traditionally
interpreted as the exchange of a Pomeron~\cite{Pomeron}. The slow rise
of the total cross section, dominated by the soft nonperturbative
strong interaction, could be described by a Pomeron trajectory with
the quantum numbers of the vacuum in Regge theory~\cite{Regge}. The
physical particles responsible for Pomeron exchange have not been
conclusively identified, but such particles can exist in Quantum
Chromodynamics (QCD) as glueballs, e.g. a $J^{PC}=2^{++}$ glueball
with a mass near 2 GeV/c$^2$~\cite{Pomeron2}. The behavior of Pomeron
exchange at low energies is not well understood because meson-exchange
processes appear and become comparable near threshold and the
applicability of Pomeron theory might be
doubtful~\cite{Donnachie}. Nonetheless, a particularly interesting and
unique way of studying the possible Pomeron exchange is
$\phi$(1020)-meson photoproduction from hadrons. In this reaction
pseudo-scalar $\pi$-meson exchange is , to first order, suppressed by
the Okubo-Zweig-Iizuka (OZI) rule in the Vector Meson Dominance model
(VDM)~\cite{VDM} because of the dominant $s \bar{s}$ quark content of
$\phi$ mesons. Furthermore, with the use of an isoscalar deuteron
target, the coupling between isovector pions and deuterons is
forbidden due to isospin
conservation~\cite{Titov_coherent,Titov_coherent2}. Accordingly, the
coherent photoproduction of $\phi$ mesons from deuterons becomes an
excellent source of information for Pomeron dynamics at low
energies~\cite{Bauer}.

In addition, coherent $\phi$-meson photoproduction provides the
opportunity to observe additional exotic processes. Possible channels
include Regge trajectories associated with particles containing
gluonic degree of freedom ({\it e.g.}, a daughter Pomeron inspired by
a scalar-type $0^{++}$ glueball~\cite{Pomeron3}, and scalar or tensor
mesons~\cite{Titov_isotopic2,Scalar_meson,Tensor_meson}). The
importance of this study is emphasized by the recent measurement of
diffractive $\phi$-meson photoproduction from protons near
threshold~\cite{LEPS_phi}. The differential cross sections
(extrapolated to zero degrees) have a local maximum around 2.0 GeV. It
was conjectured that this structure was not solely due to the
non-negligible pseudo-scalar-meson processes at low energies but is
likely signaling new dynamics beyond the Pomeron.

Besides the cross section information, the decay angular distribution
of $\phi$ mesons with respect to the photon polarization helps to
differentiate the relative contributions from natural-parity
($\sigma^N$) and unnatural-parity exchange processes
($\sigma^U$)~\cite{Schilling}. With the availability of linearly
polarized photon beams, the smallness of unnatural-parity $\pi$ and
$\eta$-exchange and hence the dominance of natural-parity processes in
the coherent production of $\phi$ mesons from deuterons can be
verified.

Recently CLAS experiment~\cite{CLAS} also measured the coherent
$\phi$-meson production from deuterons at low energies but rather at
backward angles and without the use of polarized photon beams. A total
$\phi$-N cross section at about 10 mb was determined in the framework
of vector meson dominance. The comparison of the differential cross
sections in the overlapped kinematic region between this study and the
CLAS experiment will be made later.

%%%%%%%%%%%%%%%%%
%experiment
%%%%%%%%%%%%%%%%%

\section{Experimental procedure and setup}
In this Letter, we present measurements of differential cross sections
and decay asymmetries of coherent $\phi$-meson photoproduction from
liquid deuterium near threshold in the very forward direction with
linearly polarized photons using the LEPS
spectrometer~\cite{LEPS}. Highly polarized photons were produced by
backward Compton scattering with an ultra-violet Ar laser from 8 GeV
electrons in the storage ring of SPring-8. The photon energy
($E_{\gamma}$) was determined by measuring the recoil electrons with a
tagging spectrometer event by event. A liquid deuterium target with an
effective length of 16 cm was employed. The integrated flux of the
tagged photon beams was $4.47 \times 10^{12}$ in this
analysis. Charged particles produced at the targets were detected at
forward angles with the LEPS spectrometer which consisted of a dipole
magnet, a silicon-strip vertex detector, three multiwire drift
chambers, a plastic scintillator (SC) behind the target, and a
time-of-flight (TOF) hodoscope placed downstream of the tracking
detectors. The charged particle identification was made by mass
reconstruction using both time of flight and momentum information. The
momentum resolution for 1 GeV/c particles was 6 MeV/c. The TOF
resolution was 150 psec for a typical flight path length of 4 m. The
mass resolution was 30 MeV/c$^2$ for a kaon of 1 GeV/c momentum. More
details about experimental setup are given in
Ref.~\cite{LEPS_kasym1,LEPS_kasym2}.

%%%%%%%%%%%%%%%%%
%Method and Results
%%%%%%%%%%%%%%%%%
\begin{figure}[t]
  \includegraphics[viewport=10 10 310 310,clip]{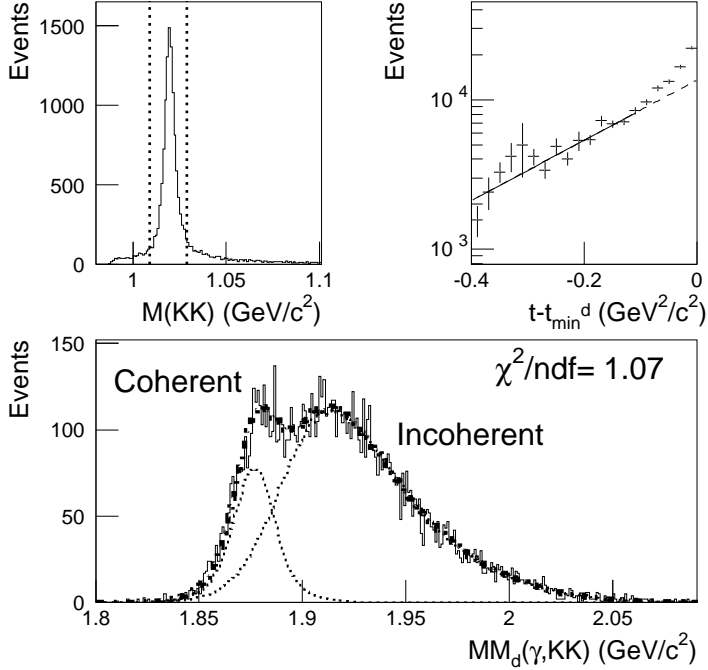}
    \caption{The distributions of invariant mass (top,left), squared
      four-momentum transfer (top,right) and missing mass for a 
      deuteron target as $\mmd$ for the events of a \KK pair
      (bottom). The two dashed lines on the distribution of invariant
      mass M(\KK) show the final cut to select the $\phi$ events. The
      $\mmd$ spectrum is shown with fit of MC-simulated coherent and
      incoherent components and the $\chi^{2}$ value of the fit. See
      the details about the lines in the plot of momentum transfer
      distribution in text.}
  \label{fig:mmdld2}
\end{figure}

The production of $\phi$ mesons was identified via the charged kaon
decay mode with the detection of $\rm{K^+}$ and $\rm{K^-}$ in the
final state. The vertex positions of \KK~ pairs were required to be
within the target boundaries. A clear signal of $\phi$ mesons was seen
in the invariant mass of \KK~pairs, M(\KK), for $E_{\gamma}$=1.5-2.4
GeV as shown in Fig.~\ref{fig:mmdld2}. A cut on the invariant mass of
$|\rm{M(\KK)}-M_{\phi}|<0.01$ GeV/c$^2$, was applied to select the
$\phi$-meson events, either through coherent or incoherent production.
The background in the selected signal region is estimated to be about
5\%-10\% by Monte Carlo simulations of coherent and incoherent
reactions and two background processes: quasi-free production of the
$\Lambda$(1520) and non-resonant \KK. We define $\tilde{t}^{\rm d}$ as
$t-t_{\rm min}^{\rm d}$ where $t$ is the squared four-momentum
transfer and $t_{\rm min}^{\rm d}$, the minimum of $t$ corresponding
to the production of $\phi$-mesons off a deuteron target at a polar
angle ($\theta$) of zero degrees.  For selected $\phi$ events in
$|\tilde{t}^{\rm d}|$$<$0.4 GeV$^2$/c$^2$, the yields as a function of
squared four-momentum transfer, $t-t_{\rm min}^{\rm d}$ (acceptance
corrected), and distribution of missing mass using a deuteron as the
rest target ($\mmd$) are also depicted in Fig.~\ref{fig:mmdld2}. Due
to the deuteron form factor and the range of acceptance, the coherent
events were observed mostly in the region of very small
$|\tilde{t}^{\rm d}|$ as illustrated by an excess of yields at
$|\tilde{t}^{\rm d}|$$<$ 0.1 GeV$^2$/c$^2$ above an exponential
extrapolation from the outer region of 0.1 $<$$|\tilde{t}^{\rm
  d}|$$<$0.4 GeV$^2$/c$^2$.

Since only the $\phi$-meson was identified in the final state, the
separation of coherent and incoherent interactions could not be
performed on a event-by-event basis. Instead, the individual yields
were disentangled by fitting the distributions of missing mass $\mmd$
where the reaction of coherent $\phi$ production from deuterons,
$\gamma d$$\rightarrow$$\phi d$, has a structure peaking at the mass
of deuterons 1.875 GeV/c$^2$, as shown in Fig.~\ref{fig:mmdld2}. This
distribution is nicely reproduced by the sum of individual ones
generated by Monte Carlo (MC) simulations. The MC simulations were
done using the GEANT3 software package~\cite{GEANT}. Geometrical
acceptance, energy and momentum resolutions and the efficiency of
detectors were included. A photon energy resolution of 10 MeV was
determined from the width of the missing mass spectra of the $\phi$
events in data from a hydrogen target.

The $\mmd$ distribution of incoherent events is affected by the Fermi
motion and the offshell effects of the target nucleons inside
deuterium. The treatment of offshell effects was found to be the
dominant source of systematic errors in this work. Two kinds of
approaches were studied for the estimate of systematic bias. In the
first approach~\cite{OFFSHELL}, the spectator nucleon was assumed to
be on-shell and the total energy of the target nucleon was determined
under the condition of Fermi momenta generated by a parameterization
with the PARIS potential~\cite{PARIS}.

Assuming isospin symmetry in the photoproduction of $\phi$ mesons from
a free nucleon, the event weighting was characterized by the measured
differential cross section from protons $d^2 \sigma$/$d
E_{\gamma}^{\rm eff} dt$~\cite{LEPS_phi} where $E_{\gamma}^{\rm eff}$
is the effective photon energy giving the same center-of-mass energy
$\sqrt{s}$ of an on-shell target nucleon at rest. The interference
between isoscalar $\eta$ and isovector $\pi$ exchange could lead to
isospin asymmetry. However the OZI effect and the smallness of the
$\eta$-N coupling strongly suppress this interference effect and thus
the isospin asymmetry of the differential cross sections in the
forward direction are expected to be small~\cite{Titov_isotopic2}.

The second approach differs from the previous one in that the degree
by which the target nucleon is offshell was randomly selected between
zero and the full scale in each event whereas the mass of target
remained unchanged after the interaction. The kinematics of the
produced $\phi$-meson were determined by the final two-body phase
space distribution. The mean values and statistic errors of the
measurements using these two approaches were averaged for the final
ones, and the difference of mean values were used for the estimation
of the systematic uncertainty.

\section{Results}
\begin{figure}[t]
  \includegraphics[viewport=10 10 310
    310,clip]{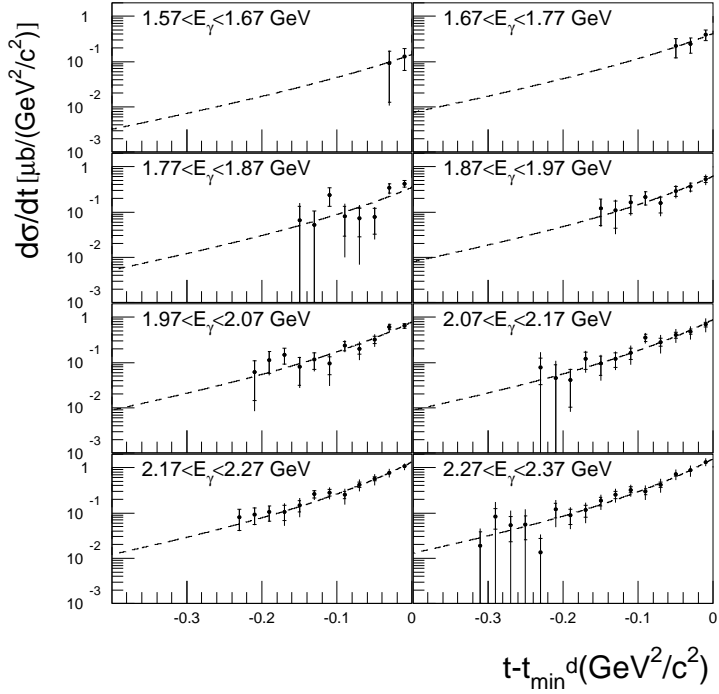}
  \caption{The differential cross sections of the coherent events $d
    \sigma / d\tilde{t}^{\rm d}$ ($\tilde{t}^{\rm d} \equiv t-t_{\rm
      min}^{\rm d}$) in various $E_{\gamma}$ bins. The smaller error
    bars represent the range of statistical errors. The dashed lines
    are the result of a fit using an exponential function multiplied
    by the deuteron form factor.}
  \label{fig:tdiscro}
\end{figure}

First we constructed the differential cross sections in the region
1.57$<$$E_{\gamma}$$<$2.37 GeV and $|\tilde{t}^{\rm d}|$$<$0.4
GeV$^2$/c$^2$. Sets of missing mass spectra of selected $\phi$ events,
in the bin sizes of 0.1 GeV for \Eg and 0.02 GeV$^2$/c$^2$ for
$\tilde{t}^{\rm d}$, were fitted with those of MC simulated coherent
and incoherent events and background processes. The distribution of
background processes overlapped with those of incoherent events. With
a proper normalization of photon beam flux, number of target atoms,
tagger efficiency, transporting efficiency and branching ratio of
charged decay of $\phi$ mesons, the differential cross sections of the
coherent events $d \sigma / d\tilde{t}^{\rm d}$ are displayed in
Fig.~\ref{fig:tdiscro}. The errors (and same below) are shown by error
bars where the smaller range is statistical errors and the whole range
is the square root sum of statistical and systematic
uncertainties. The systematic uncertainties come from the
disentanglement fit (25-30 \%), background (3\%), luminosity (5\%) and
track reconstruction efficiency (5-10\%). The fit was done with an
expression inspired by a single-scattering diagram~\cite{Glauber}:
$d\sigma/d\tilde{t}^{\rm d}$=$a e^{-b\tilde{t}^{\rm d}}$$
[F^d(\frac{1}{4}t)]^2$/$[F^d(\frac{1}{4}t_{min}^{\rm d})]^2$ where
$F^d(t)$ is the deuteron form factor and two fit parameters are $a$,
the $\gamma d$$\rightarrow$$\phi d$ cross section at $t$=$t_{min}^{\rm
  d}$, and $b$ the exponential slope. Since no strong \Eg dependence
is seen for $b$ across our energy range, a fit with a single slope
parameter $b$ was used to determine the intercept, $a$, at each \Eg
bin. The slope parameter $b$ was found to be $5.5 \pm 0.5 ({\rm stat})
\pm 0.5 ({\rm sys})$ c$^2$/GeV$^{2}$, larger than that measured in the
$\gamma p$$\rightarrow$$\phi p$ reaction $3.38 \pm 0.23$
c$^2$/GeV$^{2}$ at the same energy region~\cite{LEPS_phi}. It is noted
that the systematic error of the slope parameter is about 10\%, which
is less than the cross section uncertainties. The reason is two-fold:
for one, the main systematic errors in the cross sections $d \sigma /
d\tilde{t}^{\rm d}$ (from the uncertainties in disentangling coherent
and > incoherent) vary in a more or less coherent way across all $t$
bins; second, the slope parameter in the fitting function has a
nonlinear $t$-dependence.

\begin{figure}[t]
  \includegraphics[viewport=10 10 310 310,clip]{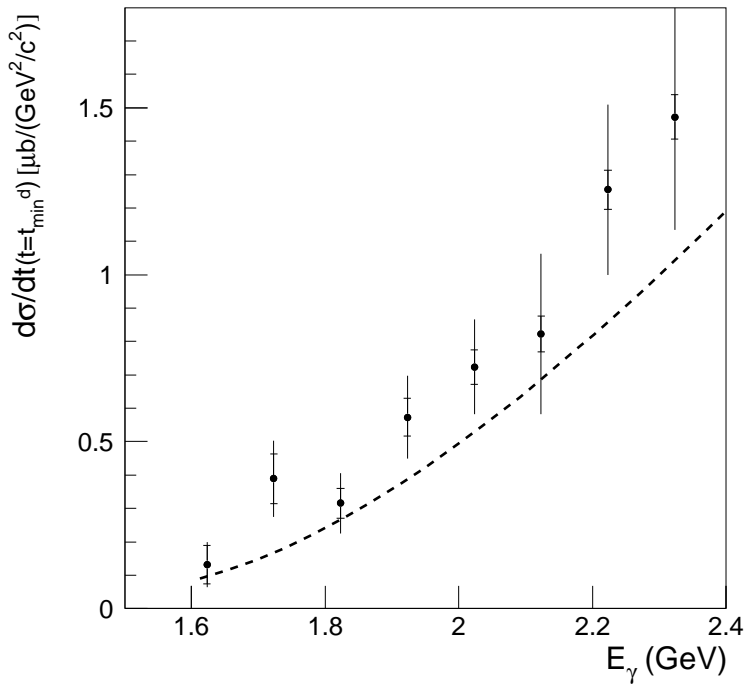}
  \caption{The fitted $d \sigma/dt$ at $t$=$t_{min}^{\rm d}$ as a
    function of photon energy. The smaller error bars represent the
    range of statistical errors. The dashed line is the predictions of
    $d \sigma^{\gamma d}/dt$ at zero degree by a model including
    Pomeron and $\eta$-exchange
    processes~\cite{Titov_coherent,Titov_coherent2}.}
  \label{fig:crossco}
\end{figure}

The $d \sigma/dt $ at $\theta$=0 determined in the coherent
$\phi$-meson events with a deuterium target as a function of photon
energy are displayed in Fig.~\ref{fig:crossco}. The energy dependence
of $d \sigma^{\gamma d}/dt$ shows a steady increase with the photon
energy. The solid line displays the calculation of $d \sigma^{\gamma
  d}/dt$ by a model taking account of Pomeron and $\eta$-exchange
processes~\cite{Titov_coherent,Titov_coherent2} and clearly the data
is under-predicted.

\begin{figure}[t]
  \includegraphics[viewport=10 10 310
    310,clip]{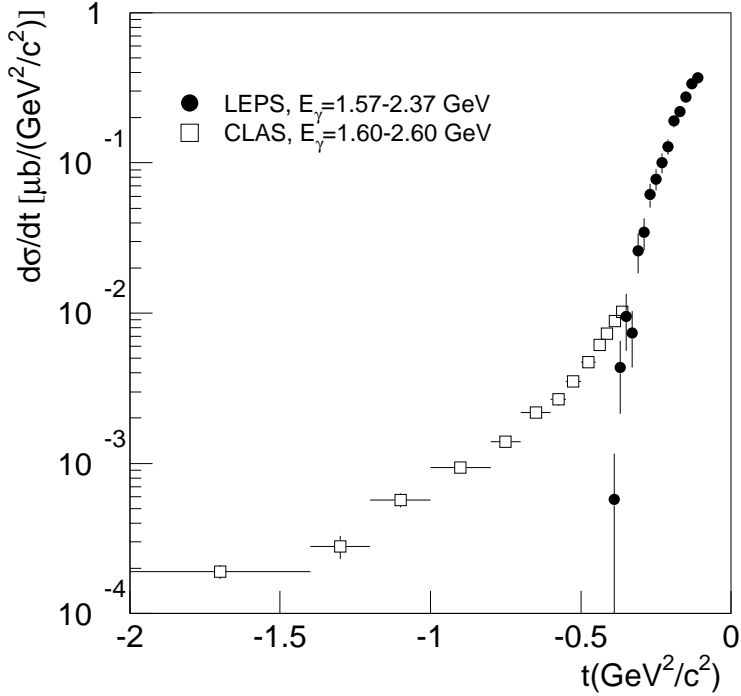}
  \caption{Comparison of differential cross sections of the coherent
    $\phi$-photoproduction from deuterons from LEPS (this study) and
    CLAS~\cite{CLAS}. Only statistical errors are displayed. The
    energy ranges for two measurements are also shown.}
  \label{fig:lepsclas}
\end{figure}

In Fig.~\ref{fig:lepsclas} we overlay two complementary measurements
of differential cross sections of coherent $\phi$-photoproduction from
deuterons at low energies: one at the forward direction for 1.57$<$$
E_{\gamma}$$<$ 2.37 GeV in the current study and the other at the
large $|t|$ region within 1.6$<$$E_{\gamma}$$<$2.6 GeV by
CLAS~\cite{CLAS}. The energy ranges for these two measurements are
mostly overlapping but it is slightly wider for CLAS. The agreement is
fairly reasonable at the overlapped $|t|$ region of these two
measurements.

The normalized decay angular distributions of W($\cos\Theta$) and
W($\Phi-\Psi$) in the Gottfried-Jackson frame for the $\phi$-produced
events were obtained in the region of $|\tilde{t}^{\rm d}|$$<$0.1
GeV$^2$/c$^2$ for 1.87$<$$E_{\gamma}$$<$2.37 GeV. Here, $\Theta$ and
$\Phi$ denote the decay polar and azimuthal angles, respectively, of
the $K^+$ in the $\phi$-meson rest frame. The y-axis of the
$\phi$-meson rest frame is perpendicular to the production plane of
the $\phi$-meson in the center-of-mass frame and the choice of z-axis
is along the momentum of the incident photon. The azimuthal angle
between the photon polarization and production plane is denoted by
$\Psi$. Events of $E_{\gamma}$$<$1.9 GeV were excluded due to
insufficient statistics in the angular bins. The polar angle
distribution W($\cos\Theta$) is consistent with (3/4)$\sin^2\Theta$,
the same as the results from protons~\cite{LEPS_phi}. This indicates
the dominance of helicity-conserving processes in $t$-channel exchange
for the photoproduction of $\phi$ mesons from deuterons.

\begin{figure}[t]
  \includegraphics[viewport=10 10 310
    310,clip]{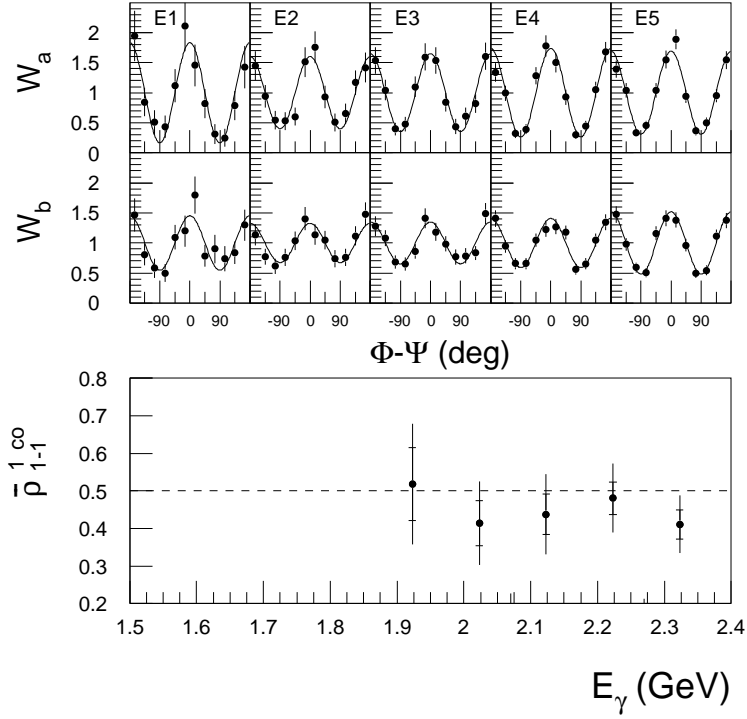}
  \caption{The decay angular distributions W$_{\rm a,
      b}$($\Phi$-$\Psi$) of \KK-pair events overlaid with the fit in
    five $E_{\gamma}$ bins of equal width and the decay asymmetry
    $\bar{\rho}^{1}_{1-1}$$^{\rm co}$ of $\gamma d$$\rightarrow$$\phi
    d$ as a function of photon energy. The subscript $a$($b$) denotes
    the events of missing mass smaller(larger) than 1.89
    GeV/c$^2$. The $E_{\gamma}$ binning starts from E1=(1.87,1.97) GeV
    and ends at E5=(2.27,2.37) GeV.}
  \label{fig:angasy}
\end{figure}

The distribution of W($\Phi-\Psi$) is parameterized as
$1+2P_{\gamma}\bar{\rho}^{1}_{1-1}\cos[2(\Phi-\Psi)]$~
\cite{Titov_spinamp}, where $P_{\gamma}$ is the degree of polarization
of the photon beams. In the case of pure helicity-conserving
amplitudes, the decay asymmetry $\bar{\rho}^{1}_{1-1}$ becomes
equivalent to half of either the parity asymmetry $P_{\sigma}$
($\equiv$
($\sigma^N$-$\sigma^U$)/($\sigma^N$+$\sigma^U$))~\cite{Titov_isotopic2}
or the decay asymmetry $\Sigma_{\phi}$ ($\equiv (\rho^{1}_{1-1} +
\rho^{1}_{11} )/(\rho^{0}_{1-1} + \rho^{0}_{11})$)
~\cite{Titov_coherent} and is 0.5($-$0.5) for pure
natural(unnatural)-parity processes. We disentangled the decay
asymmetry of coherent ($\bar{\rho}^{1}_{1-1}$$^{\rm co}$) and
incoherent ($\bar{\rho}^{1}_{1-1}$$^{\rm inco}$) interactions in the
following way. The events were divided into two, by a missing-mass
($\mmd$) cut, MM$_{div}$, of 1.89 GeV/c$^2$. Two sets of decay angular
distributions W$_{\rm a, b}$($\Phi-\Psi$) were constructed and are
shown at top of Fig.~\ref{fig:angasy}. The subscript $a$($b$) denotes
the events whose missing mass are smaller(greater) than 1.89
GeV/c$^2$. Afterwards the average decay asymmetry
$(\langle\bar{\rho}^{1}_{1-1}\rangle_{\rm a, b})$ was obtained by
fitting W$_{\rm a, b}$($\Phi-\Psi$) with
$1+2P_{\gamma}\bar{\rho}^{1}_{1-1}\cos[2(\Phi-\Psi)]$ azimuthal
distributions individually. A larger angular asymmetry was seen for
the events with smaller missing mass~\cite{PANIC06}. This is
interpreted as the difference of the mixing percentage ($R_{\rm a,
  b}$) of coherent and incoherent events distributed in the two
separate regions of $\mmd$ and their individual decay asymmetries. The
$R_{\rm a, b}$ were determined by fits of simulated missing mass
distributions and hence the individual decay asymmetry was extracted
under the assumption of linear weighting from each component,
$\langle\bar{\rho}^{1}_{1-1}\rangle_{\rm a, b}$= $R_{\rm a,
  b}\bar{\rho}^{1}_{1-1}$$^{\rm co}$+ $(1-R_{\rm a,
  b})\bar{\rho}^{1}_{1-1}$$^{\rm inco}$.

The decay asymmetries $\bar{\rho}^{1}_{1-1}$$^{\rm co}$ as a function
of photon energy at $E_{\gamma}$=1.87-2.37 GeV are shown at the bottom
of Fig.~\ref{fig:angasy}. The results are consistent in four choices
of missing mass cut for division MM$_{div}$: 1.875, 1.88, 1.89 and
1.90 GeV/c$^2$. Results using the cut of 1.89 GeV/c$^2$ are presented
because of the smallest statistical errors. Calculation of systematic
uncertainties include those in $R_{\rm a, b}$, from disentanglement
procedures (5-15\%), and the missing-mass cut for event
division MM$_{div}$ (10-20\%). As mentioned above for the slope
parameter, the systematic error of $\bar{\rho}^{1}_{1-1}$$^{\rm co}$
is less affected by the disentanglement uncertainties than those of
cross sections.

A very large decay asymmetry of $0.48 \pm 0.07 ({\rm stat}) \pm 0.10
({\rm sys})$ was observed, contrasting with a value of $0.2$ from the
proton~\cite{LEPS_phi}. Within errors, our measurement reaches the
maximum boundary corresponding to a pure natural-parity exchange
process, showing that coherent $\phi$-meson production from deuterons
is predominantly from natural-parity processes. This suggests the
absence of $\pi$-exchange, together with a negligible contribution of
$\eta$-exchange in the sector of unnatural-parity
exchanges~\cite{Titov_coherent,Titov_coherent2}.

%%%%%%%%%%%%%%%%%
%Discussion
%%%%%%%%%%%%%%%%%
\section{Discussion}

\begin{figure}[t]
  \includegraphics[viewport=10 10 310 310,clip]{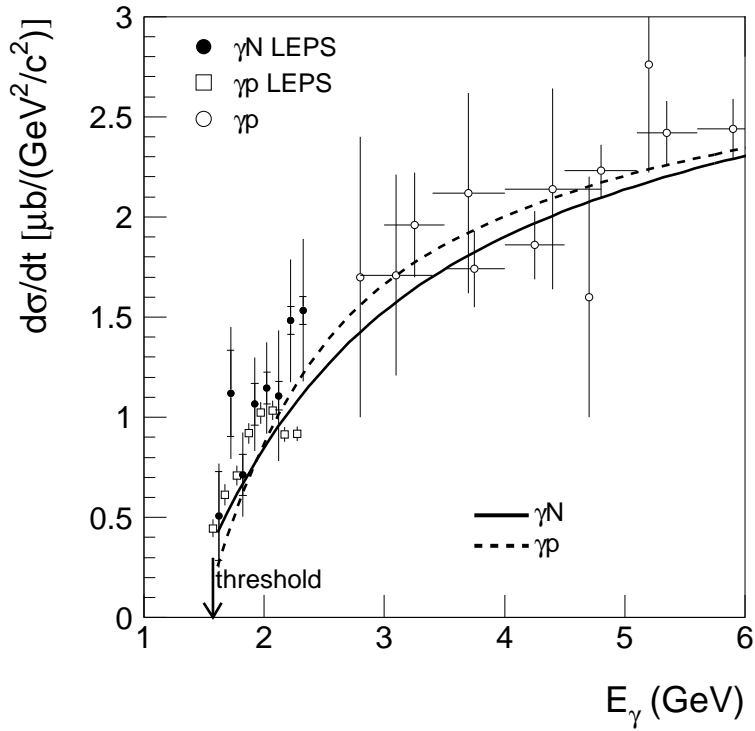}
   \caption{Data of the cross section of $\phi$-meson photoproduction
     from nucleons by isoscalar t-channel exchange processes $d
     \sigma^{\gamma N;\rm{T=0}}/dt$, deduced from coherent production
     from deuterons in this study and the existing data of $d
     \sigma^{\gamma p}/dt$ up to $E_{\gamma}$=6
     GeV~\cite{LEPS_phi,DURHAM}. All are extrapolated to
     $\theta$=0. The solid and dashed lines represent the predictions
     of $d \sigma^{\gamma N;\rm{T=0}}/dt$ and $d \sigma^{\gamma p}/dt$
     respectively by a model including Pomeron exchange and $\pi$ and
     $\eta$
     processes~\cite{Titov_coherent,Titov_coherent2,Titov_spinamp}. There
     is no contribution of isovector $\pi$-exchange in $d
     \sigma^{\gamma N; \rm{T=0}}/dt$ due to isospin conservation. The
     threshold of the $\gamma p$$\rightarrow$$\phi p$ reaction is
     labeled and the data below 2.5 GeV are shifted by $-$50 MeV for
     the clarity of display.}
   \label{fig:pomfit}
\end{figure} 

Supported by a strong dominance of natural-parity helicity-conserving
exchange processes, the $d \sigma/dt$ at $\theta$=0 for $\gamma d$$
\rightarrow$$\phi d$ are expected to reflect Pomeron exchange and the
other natural-parity exchange processes at low energies. Under the
conditions of small momentum transfer and negligible unnatural-parity
and helicity-nonconserving processes, the differential cross section
of coherent production from deuterons $d \sigma^{\gamma d}/dt$ may be
approximated as 4$S^{N}(t)$$d \sigma^{\gamma N;\rm{T=0}}/dt$ where
$S^{N}(t)$ is the natural-parity deuteron form factor and $d
\sigma^{\gamma N;\rm{T=0}}/dt$ is the cross section of $\phi$-meson
photoproduction from nucleons by isoscalar (T=0) t-channel exchange
processes~\cite{Glauber}. Depending on the choice of the total
$\phi$-N cross section between 10 and 30 mb,there is about 2-7\%
uncertainty resulting from the omission of Glauber shadowing in the
factorization approximation.

Fig.~\ref{fig:pomfit} shows the deduced $d \sigma^{\gamma
  N;\rm{T=0}}/dt$ in this study and the existing data of $d
\sigma^{\gamma p}/dt$ from threshold up to $E_{\gamma}$=6 GeV, all
extrapolated to $\theta$=0~\cite{LEPS_phi,DURHAM}, in comparison with
the corresponding theoretical
predictions~\cite{Titov_coherent,Titov_coherent2,Titov_spinamp}. The
solid line displays the calculation of $d \sigma^{\gamma
  N;\rm{T=0}}/dt$ by a model taking into account of Pomeron and $\eta$
exchange processes whereas the dashed line is for $d \sigma^{\gamma
  p}/dt$ with the inclusion of isovector $\pi$-meson exchange. It is
interesting that both $d \sigma^{\gamma N;\rm{T=0}}/dt$ and $d
\sigma^{\gamma p}/dt$ at low energies are not consistent with the
model calculation even though the data at higher energies are rather
well described. The points of $d \sigma^{\gamma N;\rm{T=0}}/dt$ which
represent the contribution by the Pomeron trajectory are mostly
under-predicted by the model. At this moment our measurement, with its
limited accuracy, cannot determine the precise energy dependence of
any specific mechanism. Nevertheless the data does hint at either a
different energy dependence for standard Pomeron exchange or the
appearance of new dynamics like a daughter Pomeron, in the
near-threshold region. For future measurements of coherent production,
it will be essential to identify the deuterons or one of the break-up
nucleons in the final state in order to reduce the systematic errors
caused by the disentanglement procedure in this work.

Information of the $\phi$-N scattering length, which is crucial for
modern QCD inspired models, could be obtained from the cross section
at zero degrees near the threshold. In Fig.~\ref{fig:pomfit}, both $d
\sigma^{\gamma N;\rm{T=0}}/dt$ and $d \sigma^{\gamma p}/dt$ appear to
be finite at the threshold of the $\gamma p$$\rightarrow$$\phi p$
reaction. It is noted that the production cross section $d\sigma$ is
zero near the threshold (because of the phase space factor), but it is
not the case for $d\sigma/dt$, which has a finite limit at the
threshold as pointed out in Ref.~\cite{Titov_kf}. A value of around
0.15 fm for the $\phi$-N scattering length is found to be consistent
with these data close to the threshold~\cite{Titov_kf}. Finally, the
observation of finite differential cross sections at threshold does
not support the use of a kinematical factor
$q_{\phi}^2$/$q_{\gamma}^2$ in the model description of
$d\sigma/dt$~\cite{Sibirtsev}, where $q_{\phi}$ and $q_{\gamma}$ are
the $\phi$-meson and photon momenta in the center-of-mass system. With
the factor $q_{\phi}^2$/$q_{\gamma}^2$, the cross section $d\sigma/dt$
is destined to be zero at threshold because $q_{\phi}$ becomes zero
and $q_{\gamma}$ remains finite.

%%%%%%%%%%%%%%%%%
%Conclusion
%%%%%%%%%%%%%%%%%
\section{Summary}
In summary we present measurements of coherent photoproduction of
$\phi$ mesons from deuterons using linearly polarized photons at
forward angles in the low energy region of $E_{\gamma}$=1.5-2.4
GeV. The cross section at $\theta$=0 shows a strong increase with
photon energy and a complete dominance of helicity-conserving
natural-parity exchange processes. The absence of unnatural-parity
isovector $\pi$-exchange, together with negligible contribution of
$\eta$-exchange is inferred. It is found that $\gamma
\langle$N$\rangle$$\rightarrow$$\phi \langle$N$\rangle$ cross sections
for isoscalar t-channel exchange at $\theta$=0 as a function of beam
energy were not consistent with the prediction of the conventional
Pomeron model. Either a modified energy dependence for the Pomeron
trajectory or additional natural-parity processes beyond Pomeron
exchange in the near-threshold region would be compatible with our
measurement. This measurement will serve as an important constraint on
the theoretical modeling of Pomeron trajectory and additional exotic
channels in the low-energy regime and help to understand the strong
coupling region of QCD. Solving the experimental challenge of directly
identifying the deuteron in the final state of forward coherent
interactions will be essential for future measurements to provide
enough accuracy to pin down the energy dependence of any specific
mechanism.

%%%%%%%%%%%%%%%%%
%Acknowledgement and bibliography
%%%%%%%%%%%%%%%%%
\section{Acknowledgments}
The authors thank the SPring-8 staff for supporting the BL33LEP beam
line and the LEPS experiment. We thank A.I. Titov for many fruitful
discussions. This research was supported in part by the Ministry of
Education, Science, Sports and Culture of Japan, by the National
Science Council of Republic of China (Taiwan), Korea Research
Foundation(KRF) and National Science Foundation (USA).

\bibliographystyle{elsart-num}

%%%%%%%%%%%%%%%%%%%%%%

%%%%%%%%%%%%%%%%%%%%%% 

\end{document}